\RequirePackage{silence}
\documentclass[fleqn,usenatbib,useAMS]{mnras} 

\usepackage{newtxtext,newtxmath}

\usepackage{graphicx}
\usepackage{amsmath}

\usepackage{amsfonts}
\usepackage{float}
\usepackage{bm}
\usepackage{ae,aecompl}
\usepackage{array}
\usepackage{soul}
\usepackage{mathtools}
\usepackage{multirow}
\usepackage[T1]{fontenc}
\usepackage[utf8]{inputenc}
\usepackage{booktabs}
\usepackage{graphicx}
\usepackage{subcaption}
\usepackage[normalem]{ulem}
\usepackage{anyfontsize}

\DeclareRobustCommand{\VAN}[3]{#2}
\let\VANthebibliography\thebibliography
\def\thebibliography{\DeclareRobustCommand{\VAN}[3]{##3}\VANthebibliography}

\DeclareRobustCommand{\appropto}{\mathrel{\vcenter{
		\offinterlineskip\halign{\hfil$##$\cr 
			\propto\cr\noalign{\kern2pt}\sim\cr\noalign{\kern-2pt}}}}}

\title[Challenges to a sharp change in $G$]{Challenges to a sharp change in $G$ as a solution to the Hubble tension} 

\author[Banik, Desmond \& Samaras]{Indranil Banik$^{1}$\thanks{E-mail: \href{mailto:indranil.banik@port.ac.uk}{indranil.banik@port.ac.uk} (Indranil Banik); \newline \hspace*{3.2em} \href{mailto:harry.desmond@port.ac.uk}{harry.desmond@port.ac.uk} (Harry Desmond)}, Harry Desmond$^{1}$ and Nick Samaras$^{2}$\vspace{10pt} \\
$^{1}$Institute of Cosmology and Gravitation, University of Portsmouth, Dennis Sciama Building, Burnaby Road, Portsmouth PO1 3FX, UK\\
$^{2}$Astronomical Institute, Faculty of Mathematics and Physics, Charles University, V Hole\v{s}ovi\v{c}k\'ach 2, CZ-180 00 Praha 8, Czech Republic}

\date{Accepted XXX. Received YYY; in original form ZZZ}

\pubyear{\the\year{}}

\begin{document}
\label{firstpage}
\pagerange{\pageref{firstpage}--\pageref{lastpage}}
\maketitle

\begin{abstract} 
It has been proposed that the gravitational constant $G$ abruptly decreased around 130~Myr ago, making Type~Ia supernovae (SNe) in the Hubble flow intrinsically brighter than those in host galaxies with Cepheid distances. This would make Hubble flow SNe more distant, causing redshifts to rise slower with distance, potentially solving the Hubble tension. We explore a wide range of unattractive consequences of this ``$G$ step model'' (GSM). We find that since the luminosities of Sun-like stars scale as approximately $G^{5.6}$, the Solar luminosity would have dropped substantially 130~Myr ago in this scenario, likely pushing Earth into a planetary glaciation. However, there was no Snowball Earth episode in the last 500~Myr. The GSM also implies that the length of a year would have abruptly increased by about 10\%, but the number of days per year has evolved broadly continuously according to geochronometry and cyclostratigraphy. The GSM would considerably alter stellar evolution, causing the Sun to have exhausted about 2/3 of its fuel supply rather than 1/2. This would make the Sun's helioseismic age exceed that of the oldest meteorite samples, but these agree excellently in practice. The expected age of the Universe also agrees well with that of the oldest Galactic stars assuming constant $G$. The GSM however implies these stars are younger, creating a lack of stars from the first 3~Gyr of cosmic history. These arguments pose significant challenges to models seeking to resolve the Hubble tension through a transition in $G$.

\end{abstract}




\begin{keywords}
    gravitation -- cosmological parameters -- cosmology: theory -- distance scale -- Sun: helioseismology -- planets and satellites: dynamical evolution and stability
\end{keywords}

\section{Introduction}
\label{Introduction}

Cosmology is currently in a crisis because the redshift $z$ of objects in the local Universe increases with their distance $r$ more steeply than expected in the standard cosmological paradigm known as $\Lambda$-Cold Dark Matter \citep*[$\Lambda$CDM;][]{Efstathiou_1990, Ostriker_Steinhardt_1995} if its parameters are calibrated using the pattern of anisotropies in the cosmic microwave background (CMB). Measurements of the CMB with the \textit{Planck} satellite imply that in a $\Lambda$CDM context, the Hubble constant $H_0 = 67.6 \pm 0.5$~km/s/Mpc \citep{Tristram_2024}, a value we denote $H_0^{\mathrm{Planck}}$ \citep{Planck_2020}. In a homogeneously expanding universe, this should be the local redshift gradient $cz'$, where $c$ is the speed of light and $z' \equiv dz/dr$. However, observations using a variety of distance indicators show that $cz'$ is about 10\% larger \citep{Riess_2024_consistency, Uddin_2024, Scolnic_2025}. This discrepancy is known as the Hubble tension \citep[for a review, see][]{Valentino_2021_problem}.

Various solutions have been proposed for the Hubble tension \citep[for a review, see][]{Valentino_2021_solutions}. One can question whether observations of the CMB really imply such a low $H_0$, which is not the case in early dark energy models \citep[][and references therein]{Poulin_2023}. If we avoid the difficulties with such approaches by assuming that $H_0 = H_0^{\mathrm{Planck}}$ \citep{Vagnozzi_2021, Vagnozzi_2023, Toda_2024} and trust the local $cz'$, we would have to question if $H_0 = cz'$, an assumption that is violated if we are living in a large local void \citep{Keenan_2013, Haslbauer_2020, Mazurenko_2024, Mazurenko_2025, Banik_2025_BAO}. We could instead alter the expansion history at late times \citep{Rezazadeh_2024}. Here, we focus on a possible scenario in which $H_0 = cz'$ but the local $cz'$ is smaller than typically quoted in the literature. Since the redshifts are spectroscopically determined, such a scenario requires that the distances to the relevant objects be larger than published. The main route to measuring the local $cz'$ involves Type~Ia supernovae (SNe) in the `Hubble flow' at redshifts of about $0.023-0.15$ or distances of about $100-600$~Mpc, where peculiar velocities should have little effect on $cz'$ \citep{Camarena_2018, Camarena_2020a, Camarena_2020b}. To calibrate the absolute SN magnitude, it is necessary to use SNe at distances $\la 40$~Mpc so that the distance to the host galaxy can be determined, typically using Cepheid variables or the tip of the red giant branch \citep[TRGB;][]{Baade_1944, Li_2024}. The period-luminosity relation of Cepheid variables \citep[the Leavitt law;][]{Leavitt_1907} and the TRGB magnitude can be calibrated in the Milky Way using geometric distances.

The Cepheid--SNe route is the most well-established method to obtain the local $cz'$. However, a problem with one or more links in this `distance ladder' could potentially invalidate the measurement. In particular, it is necessary to assume that the same Leavitt law continues to hold in galaxies too distant for geometric anchors. Likewise, SNe in host galaxies with Cepheid distances must be similar to the more distant SNe in the Hubble flow. The proposal we will focus on questions this last link in the logical chain by postulating that the gravitational constant $G$ changed abruptly $\ga 130$~Myr ago, which corresponds to distances $\ga 40$~Mpc, beyond the present range of Cepheid calibration. The $G$ step model \citep[GSM;][]{Marra_2021, Perivolaropoulos_2021, Perivolaropoulos_2024, Ruchika_2024, Ruchika_2025} exploits this gap between the outer limit of Cepheid distances to SN host galaxies and the inner limit to what can be considered the Hubble flow. Since the Leavitt law is calibrated empirically and the terrestrial value of $G$ applies to the entire Cepheid calibration zone, the GSM does not affect the Cepheid distances to SN host galaxies.

\begin{figure}
    \includegraphics[width=\columnwidth]{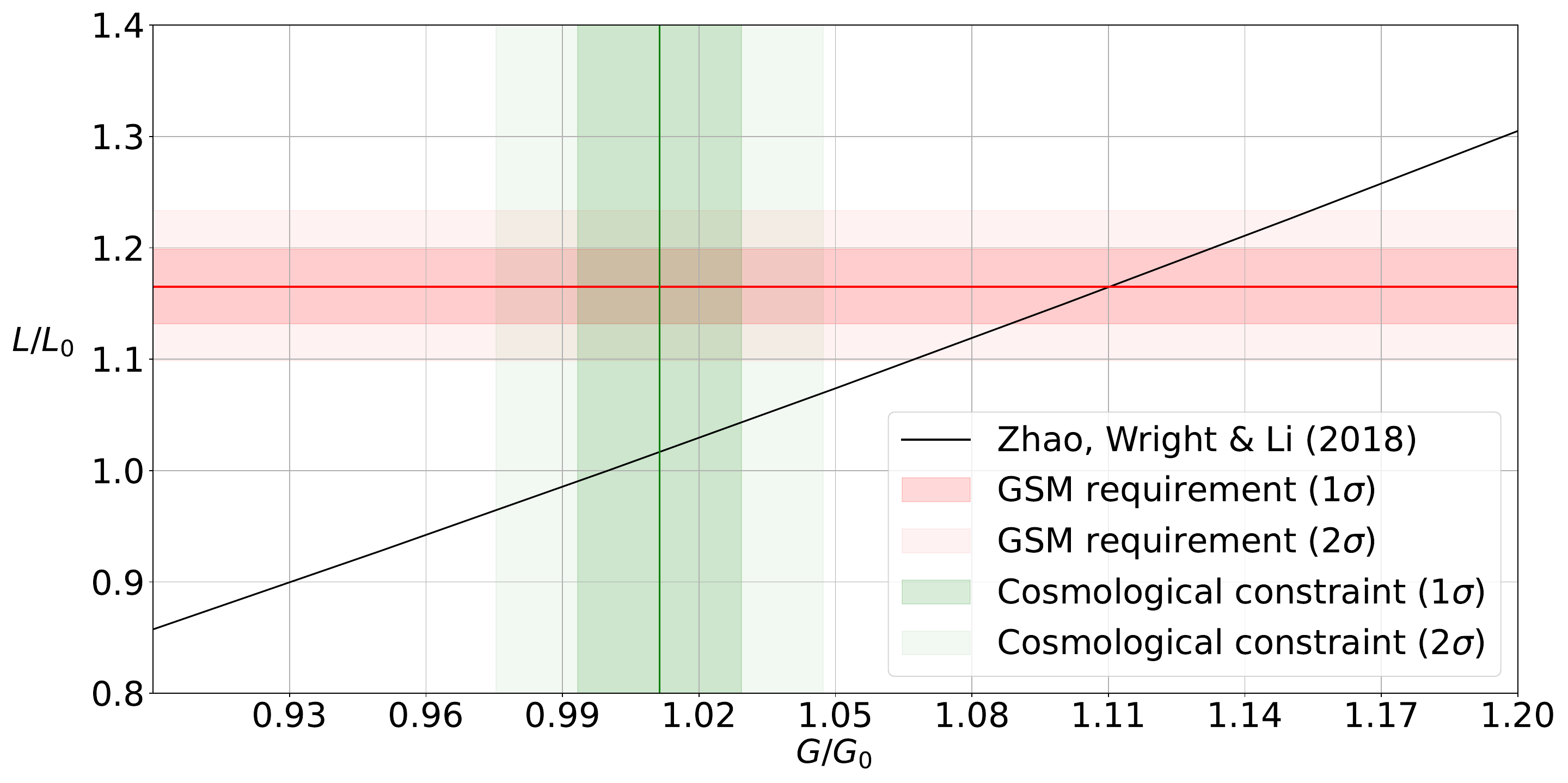}
    \caption{Comparison of constraints on and requirements of the GSM. The black line shows the dependence of the luminosity of a Type~Ia SN on $G$ in units of the terrestrial value $G_0$ \citep*[see figure~1 of][]{Zhao_2018}. This calculation accounts for changes in the shape of the light curve \citep{Wright_2018}, which impacts the standardized SN luminosity $L$ that enters cosmological analyses. Following appendix~B of \citet{Desmond_2019}, we show their $L \propto G^{1.46}$ power-law fit to the results of \citet{Wright_2018}. The horizontal red band shows the required value of $L$ in units of the standard value $L_0$ that would be needed to solve the Hubble tension by inflating the distances to SNe in the Hubble flow (Equation~\ref{Required_L_ratio}). In the GSM, these SNe were brighter than nearby SNe in host galaxies with Cepheid distances because $G$ was larger than today $\ga 130$~Myr ago. The vertical green band shows the cosmological constraint on $G$ using the latest CMB and baryon acoustic oscillation data \citep{Lamine_2025}. It is evident that the required enhancement to $G$ is in tension with cosmological constraints, even though the aim of the GSM is to preserve the \emph{Planck} cosmology. This is even before the more stringent Solar System-scale constraints described in the text.}
    \label{SN_luminosity_G}
\end{figure}

The GSM relies on the fact that $G$ affects the SN luminosity $L$ that enters into cosmological analyses \citep{Wright_2018, Zhao_2018}. The calculation is complicated somewhat by the fact that Type~Ia SNe are not standard candles but are standardizable (as indeed are Cepheids). It is well known that the peak luminosity correlates with the time required for the SN light curve to decay \citep{Phillips_1993, Phillips_1999}. Changing $G$ affects both the peak luminosity and the decay time. Cosmological analyses correct SNe for the latter (and also for the SN colour) using the Tripp formula \citep{Tripp_1998, Brout_2022}. It is therefore necessary to use the Tripp-corrected SN luminosity $L$, which has the value $L_0$ for the terrestrially measured gravitational constant of $G_0$. We use Figure~\ref{SN_luminosity_G} to show the power-law fit to the \citet{Wright_2018} relation between $L/L_0$ and $G/G_0$ \citep[$L \appropto G^{1.46}$; see appendix~B of][]{Desmond_2019}.

To explain the mismatch between $H_0^{\mathrm{Planck}}$ and the local $cz' = 73.0 \pm 1.0$~km/s/Mpc \citep{Riess_2022_cluster}, we require that
\begin{eqnarray}
    \frac{L}{L_0} ~=~ \left( \frac{cz'}{H_0^{\mathrm{Planck}}} \right)^2.
    \label{Required_L_ratio}
\end{eqnarray}
This is because the reported distances to Hubble flow SNe must be scaled by $\sqrt{L/L_0}$ to account for the proposed higher $G$ in the past. The larger distance would then reduce $cz'$ and solve the Hubble tension. The horizontal red band in Figure~\ref{SN_luminosity_G} shows the range of $L/L_0$ that would achieve this, along with the $1\sigma$ and $2\sigma$ uncertainties. The GSM requires higher $G$ prior to the transition, opposite to the change hinted at by the Tully--Fisher scaling relation \citep*{Alestas_2021}.

Since the GSM aims to preserve the \emph{Planck} cosmology, one may worry whether a higher $G$ prior to recombination would affect the CMB anisotropies. This is an important issue given that CMB observations probe well into the diffusion damping tail, which does in fact constrain the pre-transition value of $G$ relevant to cosmology \citep{Ooba_2016, Ooba_2017}. These constraints have steadily improved \citep{Ballardini_2022, Sakr_2022} and now limit the cosmologically relevant value of $G$ to within a few percent of the terrestrial value \citep{Lamine_2025}. As the cosmological observables they consider are barely sensitive to the cosmologically brief post-transition period where $G$ had its modern value, we can interpret their study as constraining the pre-transition $G$. We show their $1\sigma$ and $2\sigma$ constraints using the vertical green bands in Figure~\ref{SN_luminosity_G}. It is clear that the GSM struggles to explain early-Universe observables like the CMB consistently with the local $cz'$ measured via the Cepheid--SNe route.


In this work, we consider several other significant challenges to building an observationally viable GSM. Section~\ref{Solar_System} considers how the proposed sharp drop in $G$ would affect the Sun and the rest of the Solar System, especially the Earth. In Section~\ref{Other_stars}, we consider the impact on the evolution of other stars and how that would affect other distance indicators and measures of cosmic expansion. We summarize our results in Section~\ref{Conclusions} and conclude by considering empirically viable model extensions. In Appendix~\ref{Future_tests}, we consider a few more subtle consequences of the GSM and discuss how these might be constrained by future investigations.

\section{Impact on the Solar System}
\label{Solar_System}

The GSM has important detrimental consequences in the Solar System, which we discuss in this section. The effects on the Sun (Section~\ref{Sun}) would in turn affect Earth, both in terms of its climate (Section~\ref{Earth_climate}) and its orbit (Section~\ref{Geochronometry}). On a smaller scale, a sharp drop in $G$ would also affect the Earth--Moon system (Section~\ref{Tides}).

\subsection{Evolution of the Sun}
\label{Sun}

The nuclear reactions in the cores of stars are very sensitive to their central density and temperature, which in turn are governed by the need to maintain hydrostatic equilibrium. Detailed numerical calculations of a Sun-like star show that the Solar luminosity $L_\odot \appropto G^{5.6}$ \citep[see appendix~A of][]{Innocenti_1996}. This very steep scaling implies that if $G$ was just 5\% higher than today over the vast majority of Solar history, then $L_\odot$ would have been almost 30\% larger than with constant $G$. The more rapid evolution of the Sun would lead to a significant mismatch between its age estimated using helioseismology and the precisely known 4.567~Gyr age of the Solar System from radioactive dating of the oldest meteorite samples \citep{Connelly_2012}. However, helioseismic observations of the Sun give an age quite consistent with meteorites \citep{Guenther_1998, Bonanno_2020, Betrisey_2024}. A very substantial inconsistency is to be expected if the Sun was shining 30\% brighter than in standard theory until the very recent past, as by now it would have used up 2/3 of its fuel supply rather than 1/2. This would cause the helioseismic age of the Sun to be about 30\% larger than its actual age, so we would expect the helioseismic observations of \citet{Betrisey_2024} to give ages well above 5.5~Gyr. In reality, their figure~1 shows that the Sun has a helioseismic age $<5.1$~Gyr. This problem cannot be alleviated by postulating that $G$ has a somewhat smaller impact on $L_\odot$ than assumed above, for instance with the $L_\odot \propto G^4$ scaling proposed in equation~27 of \citet{Adams_2008} and equation~37 of \citet{Davis_2012}. Such a scaling would reduce the impact on $L_\odot$ to 20\% for a 5\% drop in $G$, yielding a predicted helioseismic age of 5.5~Gyr. Moreover, this proposed shallower scaling is only valid if opacity arises purely from Thomson scattering, which is not valid for the Sun \citep[see section~4.3 of][]{Uzan_2011}.

\subsection{Impact on the terrestrial climate}
\label{Earth_climate}

The GSM implies a significant drop in $L_\odot$, but the sharp drop in $G$ would also expand the Earth's orbital radius $r$. We will return to this issue in Section~\ref{Geochronometry}, but our main concern here is that this leads to a further drop in the Solar insolation on the Earth. Taking into account that Earth's specific angular momentum is $\sqrt{GM_\odot r}$ given its nearly circular orbit, a drop in $G$ over several years increases $r$ such that $r \propto 1/G$ in order to conserve angular momentum. Combined with the fact that $L_\odot \appropto G^{5.6}$, the Earth's blackbody temperature $T_{\oplus} \propto G^{1.9}$.\footnote{This is somewhat shallower than the earlier estimate that $T_{\oplus} \propto G^{9/4}$ \citep{Teller_1948} because that study assumed $L_\odot \propto G^7$.}

In reality, the Earth is of course not a perfect blackbody in equilibrium with the incident Solar radiation. On long timescales, its climate is maintained in the temperature range that allows liquid water oceans thanks to the carbonate-silicate cycle, which relies on the temperature dependence of weathering processes that remove the greenhouse gas CO\textsubscript{2} \citep*{Walker_1981}. However, reduced weathering would take many Myr to allow volcanic outgassing to build up enough atmospheric CO\textsubscript{2} to compensate for such a sizeable reduction in Solar insolation. It is therefore inevitable that the Earth would experience a period of significantly reduced temperature. Given that the global average $T_{\oplus} \approx 290$~K today, a $>5\%$ drop in $G$ would have reduced $T_{\oplus}$ by $\ga 10\%$ to temperatures below the freezing point of water. As water turned to ice near the polar regions, the surface would have become more reflective, further reducing the temperature. This ice-albedo feedback effect \citep{Budyko_1969, Sellers_1973} would most likely lead to a planetary glaciation lasting many Myr.

Historically, there is actually very strong evidence for several such Snowball Earth episodes \citep[for a review, see][]{Banik_2016}. However, these all took place $\ga 600$~Myr ago. This corresponds to a distance of almost 200~Mpc, where we can assume a fairly smooth distance-redshift relation. It is not possible to put the $G$ transition that far back in the past because this would lead to a clear mismatch between the luminosities of Hubble flow SNe with redshifts either side of the transition, which have a similar distance. A feature would be produced in the SN Hubble diagram, which is not observed. The only way to `hide' a substantial transition in $L_0$ (and hence in apparent magnitude at given $z$) is if it occurred sufficiently nearby that the redshift cannot be used to reliably determine the distance and thus lookback time. Peculiar velocities would `blur out' such a local transition, but not one in the Hubble flow region that starts $\ga 100$~Mpc away in standard cosmology. The transition must therefore lie closer to us, and hence cannot be matched to an observed Snowball Earth episode (see Section~\ref{Conclusions} for the possibility that the issue is instead in the lowest rung of the distance ladder). The lack of any Snowball Earth episode in the past 500~Myr implies that the proposed sharp drop in $G$ did not occur within 150~Mpc, which however is necessary for the GSM to solve the Hubble tension. This already casts significant doubt on the GSM.

An important caveat is that the response of the terrestrial climate to a sharp drop in $L_\odot$ needs to be explored in more detail to assess if a planetary glaciation could have been avoided, though models suggest that even a modest reduction in $L_\odot$ could trigger a planetary glaciation \citep{Arnscheidt_2020}. If however this can be avoided, then a more modest drop in $T_{\oplus}$ might be hidden in the geological record given other causes of temperature fluctuations and measurement uncertainties.

\subsection{Geochronometry and cyclostratigraphy}
\label{Geochronometry}

The proposed drop in $G$ would increase the length of a year (LOY) $\propto G^{-2}$. However, the length of a day (LOD) would barely change because the Earth's radius $R_{\oplus}$ would not increase much. Since angular momentum conservation applies both to Earth's orbit around the Sun and to its daily rotation on its own axis, we would require $R_{\oplus} \propto G^{-1}$ to preserve the ratio LOY/LOD. Since volume scales as radius cubed, a 5\% drop in $G$ would have to cause a 15\% drop in Earth's average density $\rho_{\oplus}$. This is unrealistic because gravitational compression has increased the uncompressed density of the Earth by only 36\%, from 4050 to 5515~kg/m\textsuperscript{3} \citep{Hughes_2006}. Thus, reducing $G$ by a mere 5\% cannot be expected to cause a 15\% drop in $\rho_{\oplus}$, especially given that Earth's interior structure and temperature were quite similar to present values 130~Myr ago as this represents only a small fraction of its age \citep{Herzberg_2010}. Another way to see this is to note that Earth rotates much slower than the Keplerian orbital velocity at its surface, so the vast majority of the support against gravity is provided by pressure gradients in nearly incompressible molten rock. As a result, reducing $G$ cannot cause Earth to expand as much as would a purely rotationally supported system. Since gravity is responsible for increasing $\rho_{\oplus}$ by 36\%, a more reasonable assumption is that a 5\% drop in $G$ would reduce $\rho_{\oplus}$ by 1.8\% and thus increase $R_{\oplus}$ by only 0.6\%, almost an order of magnitude smaller than the required 5\%. Thus, we can use the LOD as a standard clock that would not change rapidly due to a drop in $G$, unlike the LOY. This implies that the number of days per year would undergo a sharp increase on a geologically short timescale. The minimum plausible 5\% drop in $G$ would cause a 10\% increase. This corresponds to each year having only 332 days prior to the transition and the modern 365 days afterwards.

This unusual predicted behaviour can be tested using geochronometry, the idea that fossilized remains of living organisms are sensitive to familiar cycles. It is common to use tree rings to reconstruct past climate conditions and count the age of a tree in years. Of more importance for our discussion is an underwater analogue to this idea. In particular, corals rely on sunlight, making them very sensitive to day/night cycles. But their growth also varies over the year due to seasons. By studying coral growth patterns in great detail, it is possible to determine the number of days per year many hundreds of Myr in the past \citep{Wells_1963, Winter_2020}. The related technique of cyclostratigraphy exploits the related idea that deposition of sediments is also cyclic \citep{Huang_2024_geology}. Those authors report their results in terms of the LOD assuming a fixed LOY, but the actual observable is the ratio LOY/LOD, so the predicted rise in LOY would appear as a drop in their reported LOD.

\begin{figure}
    \includegraphics[width=\columnwidth]{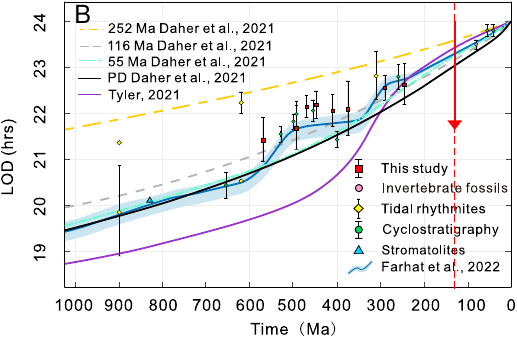}
    \caption{Evolution of the LOD from geochronometry and cyclostratigraphy as a function of lookback time, assuming a fixed LOY. The results shown here are thus inversely proportional to the observable quantity LOY/LOD, which is found from counting the number of daily cycles per annual cycle. In the GSM, the apparent LOD would have abruptly decreased by $\ga 10\%$ about 130~Myr ago (dashed red line), as indicated by the downward red arrow. The evolution at other times is due to lunar tides, as captured to some extent by the models (solid lines). Geochronometry from the last 75~Myr \citep{Winter_2020} and lunar laser ranging indicate that the LOD is presently increasing by 1~h every 150~Myr, in line with the evolution prior to the proposed transition. Adapted from figure~4b of \citet{Huang_2024_geology}.}
    \label{Huang_2024_4B}
\end{figure}

For ease of reference, we reproduce figure~4b of \citet{Huang_2024_geology} in our Figure~\ref{Huang_2024_4B}. In line with other studies, this reveals that the number of days per year continuously declined with time. This is thought to be due to days gradually getting longer due to tidal evolution of the Earth--Moon system (Section~\ref{Tides}). Tides would no doubt also operate in the GSM, but there would be an additional sharp 10\% jump superimposed on the secular evolution. However, the geological data reveal only smooth trends, with the results from $>130$~Myr ago smoothly extrapolating onto the present value. The data are therefore in tension with a 10\% jump in the number of days per year 130~Myr ago, which would appear on Figure~\ref{Huang_2024_4B} as a 10\% drop in the apparent LOD if assuming a fixed LOY. Even if we assume fairly rapid tidal evolution within the gap in the data $100-200$~Myr ago, it is clear that the apparent LOD would still have been $<24$~h 130~Myr ago, at which point it would drop 10\% to $<22$~h, as indicated by the downward red arrow. Unusually rapid tidal evolution would be required at later times to achieve agreement with our present 24~h days. This is problematic because the present rate of tidal evolution is constrained from lunar laser ranging, which is already used as a constraint on the geological models shown in Figure~\ref{Huang_2024_4B}. Given the similar positions of continents and oceans in the last 100~Myr, there is no good reason for the average rate of tidal dissipation over this period to differ much from its present rate, which if extrapolated backwards implies $>23$~h days 130~Myr ago.

Moreover, as we discuss next, the larger Earth--Moon distance after the transition and consequent reduced tidal stress on Earth would if anything cause tidal evolution to be slower rather than faster after the proposed drop in $G$, leaving the apparent LOD stuck at $\la 22$~h and thus making each year $\ga 400$~days. Therefore, although the gap in the geological record near the vertical dashed red line in Figure~\ref{Huang_2024_4B} prevents us from drawing stronger conclusions about whether a 10\% discontinuity in LOY/LOD is feasible 130~Myr ago, it does seem very unlikely that tidal evolution could reconcile such a discontinuity with available geological data.

\subsection{Impact on the Earth--Moon tidal evolution}
\label{Tides}

The proposed drop in $G$ would enlarge not only Earth's orbit around the Sun, but also the Moon's orbital radius $R$ around the Earth. Following similar arguments, the GSM implies that $R \propto 1/G$. Since the lunar tidal stress on the Earth $\propto G/R^3 \propto G^4$, a decrease in $G$ would substantially weaken oceanic tides on the Earth, which are largely caused by the Moon. On long timescales, gravitational attraction between the oceanic tidal bulges and the Moon causes it to slowly recede from Earth, a phenomenon that has been directly detected thanks to lunar laser ranging \citep{Folkner_2014}. The Moon pulls on both tidal bulges, but the bulge near it is more important due to the inverse square dependence of the gravity law. This bulge is dragged `ahead' of the Moon because Earth rotates on its axis much faster than the Moon orbits it, leading to a net tidal torque. The orbital angular momentum of the Moon ultimately comes from the rotational angular momentum of the Earth, which correspondingly slows down and thus has longer days.

The lunar recession rate and the terrestrial spindown rate must have been much higher prior to the transition. A sharp change in the terrestrial spindown rate would lead to a discontinuity in the time derivative of the number of days per year, which would evolve much more rapidly prior to the transition. However, Figure~\ref{Huang_2024_4B} shows that the spindown rate at that time (left of red line) was about the same as it is today based on lunar laser ranging, which indicates that the LOD is increasing by 2.4~ms per century or 1~h every 150~Myr \citep{Williams_2016}. It is also possible to get the present spindown rate from geochronology, which indicates that the LOD was about 0.5~h shorter than today 75~Myr ago \citep{Winter_2020}. Therefore, a discontinuity in the spindown rate 130~Myr ago is not apparent in the data over the past few hundred Myr \citep{Winter_2020, Huang_2024_geology}. All this makes it unlikely that there was a significant discontinuity 130~Myr ago both in the number of days per year and in how quickly this was evolving with time \citep[though the evolution is not completely uniform; see][]{Huang_2024_geology}.

\section{Impact on stellar probes of the distance ladder}
\label{Other_stars}

It has recently become possible to use the Fundamental Plane relation of elliptical galaxies \citep{Djorgovski_1987, Dressler_1987} to constrain the local redshift gradient anchored to the distance to the Coma Cluster \citep{Said_2024}. This lies well beyond the proposed transition, but we have to go across the transition to estimate the distance to Coma and thereby calibrate the zero-point of the Fundamental Plane distance-redshift relation. Different techniques give rather similar distances to Coma, with none suggesting that it lies $>110$~Mpc away \citep{Scolnic_2025}. However, the local $cz' = H_0^{\mathrm{Planck}}$ only if this is the case. For our discussion, the important point is the agreement between the distance to Coma as found using a variety of different distance indicators, whose absolute calibrations are generally performed closer than the proposed transition. The GSM would alter the SN distance to Coma in a manner that would solve the Hubble tension, but it seems very unlikely that the same can be said for all the other distance indicators. In particular, the luminosities of main sequence stars are much more sensitive to $G$ than the Tripp-corrected SN luminosity (Figure~\ref{SN_luminosity_G}). Because of this, it seems difficult to place Coma $>110$~Mpc away while retaining agreement between the different available distance indicators, at least if the underestimated distances are due to underestimated $G$.

Another issue with the GSM concerns the tight scaling relations evident in galaxies, especially the radial acceleration relation \citep[RAR;][]{Lelli_2017} and the related Tully--Fisher \citep{Tully_Fisher_1977} relation in spiral galaxies and the Faber--Jackson \citep{Faber_Jackson_1976} relation and Fundamental Plane in elliptical galaxies. Regardless of the underlying origin of such relations, if the $M_\star/L$ ratios of galaxies within 40~Mpc are 30\% higher than for more distant galaxies, then these scaling relations could not be as tight as they are observed to be. In particular, there would be substantial differences between samples split according to the transition distance. However, the intrinsic scatter in the RAR is very small \citep{Li_2018, Desmond_2023} and the residuals do not correlate with a wide variety of galaxy properties, including distance \citep{Stiskalek_2023}. While the latter analysis did not specifically investigate a transition at a given distance, the machine learning techniques it utilized would be capable of picking out any significant variation. This is highly problematic for the GSM because the Spitzer Photometry and Accurate Rotation Curves (SPARC) dataset in which the RAR is best measured has many precisely measured galaxies on either side of the proposed transition \citep[see figure~1 of][]{SPARC}. Moreover, the hint for a transition in $G$ in the Tully--Fisher relation identified by \citet*{Alestas_2021} is in the \emph{opposite} direction to that required by the GSM. This was not mentioned in that study due to its neglect of the effect of SN standardization on the $L-G$ relation \citep{Wright_2018}.


\subsection{Stellar ages and cosmic chronometers}
\label{CC}

As discussed in Section~\ref{Sun}, stars would be much more luminous prior to the proposed transition than in standard models of stellar evolution, which assume constant $G$. This would correspondingly shorten the lifetime of a star born near the Big Bang and just now reaching its red giant phase. As a result, the oldest stars that we observe would be $\approx 3$~Gyr younger \citep[see also][]{Davis_2012}.

The ages of the oldest stars and stellar populations in the Galactic disc and halo \citep{Cimatti_2023} agree quite well with the age of the Universe in the \emph{Planck} cosmology assumed by the GSM \citep[e.g.,][]{Banik_2024, Mazurenko_2025}. This agreement applies to standard calculations of the stellar ages. If instead the ages are recalculated assuming higher $G$ over nearly the entire lifetime of each star, then there would be an $\approx 3$~Gyr mismatch \citep{Innocenti_1996}. It would be quite unusual if no stars were identifiable from the first 3~Gyr, especially when considering that several galaxies have already been discovered by the \emph{James Webb Space Telescope} (\emph{JWST}) at $z > 14$ \citep{Carniani_2024}. In the \emph{Planck} cosmology assumed by the GSM, this corresponds to only 0.3~Gyr after the Big Bang, which may be faster than expected in the $\Lambda$CDM model \citep{Haslbauer_2022_JWST}. A more severe version of this age gap problem applies to the covarying coupling constants model, which predicts that the universe is 26.7~Gyr old \citep{Gupta_2023}.

It is also possible to constrain the expansion history using cosmic chronometers \citep[CCs;][]{Moresco_2018, Moresco_2020}. The basic idea is to get the time elapsed between two redshifts by considering `red and dead' quiescent elliptical galaxies that presumably formed their stars at very early times. The stellar population of such a galaxy would evolve passively, with only the less massive stars remaining at later times. This would change the relative strengths of spectral lines, allowing a determination of the relative age between galaxies at the two considered redshifts. Of particular importance to the CC technique is the D4000 break at 400~nm \citep{Moresco_2024}. Calculations would need to be conducted of how this feature evolves with time at higher $G$. More generally, we would need to obtain a revised relation between the colour of a star and its lifetime. Although higher $G$ reduces the lifetime at fixed mass, it is possible that the colour also changes in such a way as to largely move stars along the standard colour-lifetime relation. Even so, it seems unlikely that such cancellation of different effects would be precise enough to explain the excellent agreement between the \emph{Planck} background cosmology and that reconstructed using the CC technique \citep{Gomez_2018, Cogato_2024, Guo_2025}. The latter authors found that $H_0 = H_0^{\mathrm{Planck}}$ to within 1~km/s/Mpc, which is also their stated uncertainty. It is difficult to understand how this agreement can be preserved in the GSM, even if stellar luminosities scale with $G$ somewhat less steeply than expected theoretically \citep[$G^{5.6}$; see][]{Innocenti_1996}. It therefore seems very problematic that a change in $G$ would necessarily apply to both Hubble flow SNe and main sequence stars. This serious problem with the GSM could potentially be avoided in a more fundamental theory that yields a sharp drop in $G$ or in models with a screening mechanism, which we discuss in the next section.

\section{Summary and model extensions}
\label{Conclusions}

The GSM hypothesis postulates that due to a sharp reduction in $G$ by about 10\% in the last $\approx 50-300$~Myr, Type~Ia SNe in the Hubble flow have a higher luminosity than SNe in host galaxies near enough to have their distance measured using the Leavitt law of Cepheid variables \citep{Marra_2021, Perivolaropoulos_2021, Perivolaropoulos_2024, Ruchika_2024, Ruchika_2025}. As a result, distances to Hubble flow SNe would have to be revised upwards, reducing the locally measured $cz'$ and plausibly reconciling it with the $\Lambda$CDM prediction based on the \emph{Planck} cosmology, which is chosen to fit the CMB anisotropies without regard to the local $cz'$.

We find that the GSM has several unintended consequences because $G$ affects a plethora of other astrophysical observables besides the luminosity of Type~Ia SNe. The main issues we identified with the GSM are as follows:
\begin{enumerate}
    \item The Solar luminosity $L_\odot \appropto G^{5.6}$ \citep{Innocenti_1996}. If $G$ were higher for most of the Sun's history, it would have consumed much more of its fuel in the precisely known time since it formed \citep{Connelly_2012}. This would inflate its helioseismic age to well over $5.5$~Gyr, which is not the case observationally \citep{Betrisey_2024}.

    \item A sharp drop in $G$ and thus $L_\odot$ (combined with the resulting expansion of Earth's orbit) would likely have plunged the Earth into a planetary glaciation due to the ice-albedo feedback effect \citep{Budyko_1969, Sellers_1973}. Although there have been several Snowball Earth episodes, none occurred in the time period relevant to the GSM \citep[][and references therein]{Banik_2016}.

    \item A reduction in $G$ would cause Earth's orbit to expand, increasing the LOY. However, the LOD would not change in like manner because the Earth is almost exclusively supported against gravity by chemical forces rather than rotation. Compression of the Earth by gravity only raises its density by 36\% \citep{Hughes_2006}, so a reduction in $G$ by 5\% cannot be expected to reduce its overall density by 15\%, which however would be needed to increase its radius by 5\% in order to increase the LOD in line with the LOY. As a result, the number of days per year (LOY/LOD) would sharply increase by $\ga 10\%$. It is possible to constrain this using geochronometry and cyclostratigraphy (Section~\ref{Geochronometry}), for instance using corals that are sensitive to both daily and seasonal cycles \citep{Wells_1963}. There is no indication of such a rapid change in the number of days per year \citep{Winter_2020, Huang_2024_geology}. Instead, extrapolating trends in the data prior to the $G$ transition gives a value close to the present-day value of 365.25. A gap in the geological data $100-200$~Myr ago prevents a clearer assessment of whether the sharp change predicted by the GSM actually occurred, though quite unusual tidal evolution of the Earth--Moon system would be required in this period to mask the effects predicted by the GSM (Figure~\ref{Huang_2024_4B}).

    \item The expansion of gravitationally bound orbits would also apply to the Moon. A rapid increase in the Moon's distance from Earth combined with a reduced $G$ would sharply decrease the amplitude of Earth's oceanic tidal bulges. Since gravitational interactions between these bulges and the Moon cause the Earth's spin to slow down, there would be a discontinuity in the time derivative of the number of days per year, which must have evolved much more rapidly prior to the transition. There is again no evidence of this, with lunar laser ranging and geochronometry within the last 100~Myr indicating that the LOD is increasing by about 1~h every 150~Myr, in line with the evolution prior to the transition (Figure~\ref{Huang_2024_4B}). This suggests that the distance to the Moon did not jump by 10\% on a geologically short timescale in the last few hundred Myr.

    \item The local $cz'$ can be obtained using several other techniques besides the traditional Cepheid--SNe route \citep[e.g.][and references therein]{Riess_2024}. Since these give quite similar results, it is unlikely that the evolution of $G$ can be chosen to alter all the distances in step with each other. In particular, the GSM implies that stars were substantially brighter prior to the proposed transition. This would lead to a clear mismatch between distance techniques that rely on stellar luminosities and those that do not. For instance, we would expect a discrepancy between redshift and surface brightness fluctuation (SBF) distances, which are sensitive to the luminosity-weighted mean stellar luminosity \citep{Cantiello_2023}. A large difference between the $M_\star/L$ ratios of galaxies within and beyond the proposed transition would also introduce a large mismatch in galaxy scaling relations like the Tully--Fisher relation and Fundamental Plane. It is particularly difficult to understand how galaxy samples that extend both sides of any plausible transition radius follow such a tight RAR \citep{Desmond_2023}, especially given that a model-agnostic machine learning-based approach to checking if residuals from the RAR correlate with heliocentric distance found no discernible correlation \citep{Stiskalek_2023}.

    \item The GSM implies substantially faster stellar evolution prior to the transition, and thus over the vast majority of cosmic history. This would cause the stellar populations of quiescent galaxies to evolve much more between any two redshifts, which would have significant effects on reconstructions of the expansion history using CCs. Higher $G$ would however also affect other aspects of a star at any fixed mass that might be used to estimate its age, for instance its colour and the strength of the D4000 feature in its spectrum. Even so, it is unlikely that these effects would balance so precisely that the CC technique would recover $H_0 = H_0^{\mathrm{Planck}}$ to within 1~km/s/Mpc, as occurs in practice \citep{Cogato_2024, Guo_2025}. Moreover, the proposed higher $G$ in the past would not apply to current observations of the oldest stars and stellar populations in the Galactic disc and halo given their proximity. Estimating their ages assuming constant $G$ yields a cosmic age in good agreement with the \emph{Planck} cosmology \citep{Cimatti_2023, Banik_2024, Mazurenko_2025, Valcin_2025}. The GSM would force a substantial downward revision to these ages because these stars would have experienced higher $G$ for most of their history, leading to an age gap of $\approx 3$~Gyr in the early universe lacking any known stellar counterpart.
\end{enumerate}

Our consideration of these issues suggests that the GSM faces significant difficulties when additional constraints are imposed on the evolution of $G$ over the past $\approx 300$~Myr, the light travel time to the nearest Hubble flow SNe. Moreover, the local redshift gradient can be obtained in several ways that do not rely on the traditional Cepheid--SNe route, yet give similar results. We therefore argue that the GSM as it currently stands faces significant challenges that more realistic models would need to overcome in order to be viable.

It is important to note three caveats to our conclusions. The first is that the theoretical modelling we employ is somewhat limited, e.g., the helioseismic bound in Section~\ref{Sun} is subject to the caveat that we did not do a full recalibration of the model for the Sun with a time-varying $G$. This may lead to uncharacterized theoretical uncertainties. Secondly, the tests we employ do not incorporate full goodness-of-fit comparison or hypothesis testing, and hence may be considered statistically primitive according to rigorous astrophysical standards. This may introduce additional systematic uncertainty. We therefore look to future data and analyses to shore up our conclusions. Finally, the GSM is currently only a phenomenological model. A fundamental theory that produces this phenomenology -- perhaps through the dynamics of new gravitational fields -- may produce somewhat different predictions that fare better against the observables we consider. Our constraints may therefore be interpreted as guidelines and challenges for model-builders.

The GSM focuses on the possibility that the local $cz'$ has been overestimated because of a mismatch between the luminosities of SNe in the Hubble flow and those in more nearby host galaxies with an independent Cepheid or TRGB calibration. The difficulties with the GSM lead us to consider if the problem might lie rather with the first rung, which assumes that the Leavitt law and TRGB magnitude are the same between geometric anchor galaxies and SN host galaxies. One way to violate this assumption would be through a rapid evolution in $G$, which would need to occur even more recently than in the GSM. This would if anything exacerbate the already significant problems we have identified with the GSM. On the other hand, comparison of Cepheid and SNe distances within 30~Mpc provides some hints for a change in $G$ within the last 100~Myr based on a distance-dependent transition in the SN~Ia absolute magnitude, exploiting the different way in which $G$ affects SNe and Cepheid variables \citep{Perivolaropoulos_2021, Ruchika_2024}. In particular, \citet{Ruchika_2025} found evidence for a model in which $G$ decreased by 4\% 80~Myr ago. However, their analysis requires too strong a $G$ dependence of the standardized SN luminosity. Those authors parametrize the dependence in the form $L \propto G^{-3n/2}$, with the theoretical expectation being $n = -0.97$ \citep{Wright_2018}. However, the joint inference on $n$ and $H_0$ is inconsistent with $n = -0.97$ and $cz' = H_0^{\mathrm{Planck}}$ at well over $2\sigma$ confidence (see their figure~5). Indeed, their most preferred model has $n = -4.24 \pm 1.06$, which is $>3\sigma$ discrepant with the theoretical expectation. Some of their models can largely resolve the Hubble tension if we allow $n \approx -2$, but this would require significant unknown errors in the calculations of \citet{Wright_2018}. Moreover, a sharp change in $G$ at the reduced distance of 25~Mpc would exacerbate the problem discussed earlier concerning the tightness of the RAR and lack of residual correlations with heliocentric distance \citep{Stiskalek_2023}, as 25~Mpc is closer than 40~Mpc to the median distance of the SPARC sample.

Most if not all of these issues can however be resolved by replacing the GSM with a more physically motivated model in which $G$ effectively correlates with environmental density due to screening, naturally hiding the modification within the Galaxy. This may shield the model from anomalous effects in stars and their planetary systems $-$ from which most of the above constraints derive $-$ while still affecting the distance ladder enough to reduce the local $cz'$ down to $H_0^{\mathrm{Planck}}$. Indeed, screening mechanisms are generic in modified gravity theories~\citep[for recent reviews, see][]{Baker_2021, Brax_2021}.

A model along these lines was developed in \citet{Desmond_2019}, where it was proposed that the Leavitt law in the anchor galaxies NGC~4258, the Milky Way, and the Large Magellanic Cloud is different to that in SN host galaxies due to screening effectively correlating $G$ with one of various measures of environmental density, including local dark matter density \citep{Sakstein_2019}. Such a model does not modify SNe themselves, but there is an effective difference between the period--luminosity relations of Cepheids in galaxies with geometric anchors versus SN host galaxies, which are used to calibrate the SN absolute magnitude. This alters the distances to all Hubble flow SNe by the same factor, with corresponding impact on the local $cz'$ given the lack of any impact on $z$. Imposing the constraint on $G$ from Cepheid--TRGB consistency, this allows the SH0ES Hubble tension to be reduced to $\approx 2.5\sigma$ \citep{Desmond_2019}, while that for the TRGB-calibrated distance ladder is completely resolved \citep{Desmond_2020}. Although it is unclear if this remains the case with the latest measurements from the \emph{JWST} \citep{Freedman_2024, Riess_2024_consistency, Jensen_2025}, the screening model appears more attractive than the GSM due to its natural circumvention of the constraints we describe here as well as its far stronger theoretical motivation. Variants of this model are studied in \citet{Hogas_2023a, Hogas_2023b}.

\section*{Acknowledgements}

IB and HD are supported by Royal Society University Research Fellowship grant 211046. NS is supported by the AKTION grant number MPC-2023-07755, the Charles University Grants Agency (GAUK) grant number 94224, and the Deutscher Akademischer Austauschdienst Bonn-Prague exchange program. The authors thank Earl Bellinger, Charles Dalang, Hugo Moreira, and Jeremy Sakstein for helpful discussions. They also thank the referee for comments which improved this contribution.

\section*{Data Availability}

No new data were created or analysed in this contribution.

\bibliographystyle{mnras}
\bibliography{GSR_bbl}

\begin{appendix}

\section{Additional consequences of the GSM and future tests}
\label{Future_tests}

In this section, we collect a few more qualitative or speculative consequences of the GSM that could be used to construct stringent tests with future data and analysis.

The distance ladder can be calibrated with many different types of star, whose luminosities exhibit a variety of dependencies on the value of $G$. Thus, distance indicators calibrated to match each other within 40~Mpc would likely become discrepant beyond the proposed transition. For instance, Cepheid variable pulsations respond differently to main sequence stars \citep{Jain_2013, Sakstein_2013}, whose average luminosity is essentially what the SBF technique measures \citep{Cantiello_2023}. The TRGB magnitude \citep{Anand_2022, Anand_2024} responds still differently \citep{Sakstein_2019}. Geometric distances to megamasers would not be much affected \citep{Pesce_2020}; as all but one of the megamasers considered by those authors are beyond the GSM transition, further constraints could be obtained by comparing megamaser distances to those obtained in ways that are sensitive to $G$.

The GSM would also affect the propagation of gravitational waves \citep[GWs;][]{LIGO_2016}. It has been argued that GWs would propagate continuously, even during the period in which $G$ varies rapidly \citep{Paraskevas_2023}. Nonetheless, it seems inevitable that the GSM would have different effects on distances based on stellar populations and those obtained using GWs as standard sirens. Further progress is possible using GWs with an electromagnetic counterpart, with one such instance having been reported so far \citep[GW170817; see][]{LIGO_Virgo_2017}. Its distance of 40~Mpc is right where the transition is proposed to have occurred. In the future, it should be possible to find similar events either side of the transition. This might eventually allow a determination of the maximum mass of neutron stars, which is sensitive to $G$ and would therefore differ either side of the transition \citep{Reyes_2024}. The GSM would also lead to a significant change in the gravitational binding energy of any compact object, with neutron stars particularly strongly affected \citep{Goldman_2024}. Moreover, the entropy of a black hole $\propto G$, so a sharp drop in $G$ seems incompatible with the second law of thermodynamics \citep{Putten_2024}.

While the GSM would undoubtedly have considerable impact on the Sun (Section~\ref{Sun}), it can be easier to model slightly less massive stars, yielding less model-dependent constraints on possible evolution of $G$. \citet{Bellinger_2019} use asteroseismology of the star KIC~7970740 to constrain a model in which $G$ gradually changes as a power-law in time since the Big Bang. The impact would be more severe than the gradual changes over a Hubble time considered by those authors if $G$ were 5\% larger than today for nearly the entire 11~Gyr history of KIC~7970740. It would be interesting to constrain the GSM with asteroseismology of this or other stars. However, the age would have to be inferred alongside the other parameters. The Sun offers an advantage in this respect because its age is known from radioactive dating of rock samples.

The GSM would cause bound, nearly circular orbits to expand $\propto 1/G$, an increase which would also affect the Moon (Section~\ref{Tides}). It is unfortunately still difficult to detect the monthly cycles caused by the changing angle between the Moon and Sun causing cyclic variations in the tidal range. Future work may help to identify these cycles more reliably and thus better constrain the number of days between full Moons, which in the GSM would reveal a discontinuous behaviour.

If the proposed reduction in $G$ occurred on a short enough timescale, any bound orbits would be somewhat destabilized. A sudden drop in the equilibrium circular velocity would cause an object initially on a circular orbit to find itself on a wider elliptical orbit. In the case of the Earth, this situation would arise if the transition occurred over $\ll 1$~yr. Its eccentricity would be `pumped' by a sharp change in $G$. This is disfavoured by the fact that Earth's orbital eccentricity is only 0.017 and that of Venus is only 0.007.

There would also be consequences elsewhere in the Solar System, including on the giant planets. In general, the whole Solar System would be somewhat destabilized. This would increase the likelihood of giant impacts on the Earth. Indeed, this very possibility has been related to the extinction of the dinosaurs and many other species in the Cretaceous--Paleogene extinction \citep{Perivolaropoulos_2022}, which occurred 66~Myr ago \citep{Renne_2013}. This corresponds to a distance of only 20~Mpc, but there are Type~Ia SNe with Cepheid calibration out to $\approx 40$~Mpc \citep[e.g.,][]{Riess_2022_comprehensive}. Even so, recent studies suggest that the GSM can be reconciled with a transition at a lookback time corresponding to only 20~Mpc \citep{Ruchika_2024, Ruchika_2025}. It therefore remains possible that the GSM is associated with the Cretaceous--Paleogene extinction event, though this is disfavoured by geochronometric constraints on the number of days per year a few Myr prior to this very clear geological boundary \citep[see figure~7 of][]{Winter_2020}. It is not presently known whether the asteroid impact largely responsible for it was an isolated incident or part of a more general increase in giant impacts at that time, as would be expected in the GSM.

The eccentricity pumping effect discussed above could be mitigated if the transition occurred adiabatically, with $G$ remaining roughly constant on an orbital timescale. This is certainly possible from a distance ladder perspective: the GSM would still work if the transition took several kyr or even a few Myr. However, Galactocentric orbits are necessarily much longer than would be available for the proposed transition. This could lead to unusual effects on orbiting stars and gas, possibly leading to an enhanced star formation rate as gas on an initially circular orbit is driven on to a more eccentric orbit. The orbital timescales are even longer for tidal streams due to their larger Galactocentric distance. A sharp drop in $G$ would reduce the Galactic gravity on any satellite, not only directly but also indirectly by allowing the Galactic dark matter halo to expand, leading to less enclosed mass within the orbit of the satellite. This reduction in gravity would lead to a discontinuous curvature of the satellite's trajectory, which might be detectable in the Sagittarius tidal stream \citep{Ibata_2001, Newberg_2002}. Fainter tidal streams might be better suited to finding or ruling out the expected signature because the satellite would have a lower internal velocity dispersion, leading to a thinner and more clearly defined stream.

Finally, returning to the smaller scale of the Earth, a change in $G$ would affect how it maintains hydrostatic equilibrium. The proposed reduction in $G$ would cause it to expand slightly, as overlying layers of rock exert less weight. This may lead to unusual tectonic effects, possibly triggering earthquakes and volcanism. Life on the Earth would have to respond to these effects, and also more directly to the reduced Solar radiation and surface gravity. Since complex life was widespread by the time of the proposed transition, it may have had interesting consequences for especially large land animals like dinosaurs $-$ unless the transition is associated with their demise. This of course depends on the duration of the transition, which we have generally assumed would occur over $\gg 1$~yr. Life would undoubtedly adapt to a more gradual change in $G$ over several kyr or Myr, though there might be interesting evolutionary consequences.

\end{appendix}

\bsp 
\label{lastpage}
\end{document}